\documentstyle[12pt,epsf]{article}
\newcommand{\SS}{{\cal S}}
\newcommand{\VV}{{\cal V}}

\newcommand{\OO}{{\cal O}}

\newcommand{\be}{\begin{equation}}
\newcommand{\ee}{\end{equation}}
\newcommand{\ben}{\begin{eqnarray}\displaystyle}
\newcommand{\een}{\end{eqnarray}}
\newcommand{\refb}[1]{(\ref{#1})}
\newcommand{\p}{\partial}

\begin{document}

\begin{flushright}
hep-th/0009191\\
MRI-P-000905
\end{flushright}

\vskip 3.5cm

\begin{center}
{\Large \bf Normalisation of the Background Independent}\\
\medskip
{\Large \bf  Open String Field Theory Action}

\vspace*{6.0ex}

{\large \rm Debashis Ghoshal\footnote{{\tt ghoshal@mri.ernet.in}} 
and Ashoke Sen\footnote{{\tt ashoke.sen@cern.ch}, {\tt sen@mri.ernet.in}}}

\vspace*{1.5ex}

{\large \it Mehta Research Institute}\\ 
{\large \it of Mathematics \&\ Mathematical Physics}\\
{\large \it  Chhatnag Road, Jhusi}\\
{\large\it Allahabad 211019, India}

\vspace*{4.5ex}

{\bf Abstract}

\begin{quote}
It has been shown recently that the background independent open string
field theory provides an exact description of the tachyon condensation on
unstable D-branes of bosonic string theory. In this analysis the overall
normalisation of the action was chosen so that it reproduces the
conjectured relations involving tachyon condensation. In this paper we fix
this normalisation by comparing the on-shell three tachyon amplitude
computed from the background independent open string field theory with
the same amplitude computed from the cubic open string field theory, 
which in turn agrees with the result of the first quantised theory. We 
find that this normalisation factor is in precise agreement with the one
required for verifying the conjectured properties of the tachyon
potential.
\end{quote}
\end{center}

\newpage


\baselineskip=17pt

The 26-dimensional bosonic string theory contains D-$p$-branes for all
$p$. Each of these D-$p$-branes has a tachyonic mode. It has been
conjectured\cite{9902105} that there is a local minimum of the tachyon
potential which describes the closed string vacuum without any D-brane. At
this
minimum the negative contribution from the tachyon potential exactly
cancels the tension of the D-brane. Further, it has been conjectured that
a codimension $q$ lump solution on the D-$p$-brane represents a 
D-$(p-q)$-brane in the same theory. Support for these conjectures comes 
from the analysis of the world-sheet theory\cite{WORLD,9902105}, cubic
open string field theory (COSFiT)\cite{SFTAN}, noncommutative limit 
of the effective field theory of the tachyon\cite{NONCOM}, as well as 
various toy models of tachyon condensation\cite{padic,TOY}.

A different open string field theory that is (formally) background
independent was proposed and developed in 
Refs.\cite{9208027,9210065,9303067,9303143,9311177}.
Recently it has been pointed out\cite{0009103,0009148} that
this string field theory can provide an exact
verification of these conjectures. A general field configuration 
in background independent open string field theory (BIOSFiT) is 
associated with a boundary operator of ghost number 1 in
the world-sheet field theory of matter and ghost system. 
We shall take the world sheet to be a disc of unit radius with
flat metric on it and
work in the convention $\alpha'=1$. If $\{\OO_I\}$
denotes a complete set of boundary vertex operators 
of ghost number 1,
we can expand a general operator $\OO$ of ghost number 1 as
\be \label{e2}
\OO=\sum_I\lambda^I \OO_I \, .
\ee
We shall restrict to operators of the form $\OO=c\VV=\sum_\alpha
\lambda^\alpha c
\VV_\alpha$,
where $c$ is the ghost field and $\VV=\sum_\alpha\lambda^\alpha
\VV_\alpha$
is
a
boundary operator in the matter theory. In this case, a string field
theory
configuration associated with the operator $c\VV$ is described by the
world-sheet action 
\be \label{e1}
\SS_{Bulk} + \int_0^{2\pi}{d\theta\over
2\pi} \VV(\theta)\, ,
\ee
where the angle $\theta$ parameterises the boundary of the disc, 
and $\SS_{Bulk}$ denotes the bulk world-sheet action corresponding to
the closed string background. We shall consider a trivial background
in flat space, therefore
$\SS_{Bulk}$ describes the CFT of 26 free scalar fields $X^\mu$ and
the $(b,c)$ ghost system.
For such configurations, the string field theory action
$S_{BI}(\lambda^\alpha)$ is
obtained as a solution of the equation:
\be \label{e3}
{\delta S_{BI}\over \delta \lambda^\alpha} = {K\over 2} \int {d\theta\over
2\pi} \int
{d\theta'\over 2\pi} \left\langle \OO_\alpha(\theta) \{Q_B,
\OO(\theta')\}\right\rangle_\VV\, ,
\ee
where $\langle\,\cdots\,\rangle_\VV$ denotes correlation
function in
the world-sheet field theory described by the action
\refb{e1}. $Q_B$ is the BRST charge and $K$ is a normalisation constant 
to be fixed later. In the following, we
shall also use the correlation function $\langle\,\cdots\,\rangle$
in the absence of the boundary term in \refb{e1}. Notice that 
eqn.(\ref{e3}) determines the action upto an additive constant. 
However, since we shall always compute the difference between the 
values of the action for two configurations, this
ambiguity will not affect our analysis. The subscript $BI$ in
eqn.\refb{e3}
stands for background independent open string field theory.

A special class of string field configurations corresponding to operators 
of the form $\OO=c\VV$ with: 
\be \label{e4} 
\VV = a + \sum_i u^i (X^i)^2\, ,
\ee
was analysed in Refs.\cite{9210065,0009148}.
In this case the action can be computed {\em exactly} since the 
worldsheet theory remains free. 
The resulting action has an unstable extremum at $(a=0, u^i=0)$
corresponding to the
original D-brane. In addition, it has several other extrema with
the following properties:  
\begin{enumerate} 
\item There is an extremum at
$(a=\infty, u^i=0)$. The difference in energy density between the original
configuration $(a=0, u^i=0)$ and this extremum is
$K$\cite{0009148}.\footnote{Throughout this paper we shall be using the
convention that when $\VV=0$, the partition function of the matter
conformal theory on the unit disk is equal to the volume $\Omega$ of the
D-brane
world-volume. With this convention, $S_{BI}(a, u^i=0) = \Omega K (1+a)
e^{-a}
+C$
where $C$ is an additive constant. This differs somewhat from the
convention used in ref.\cite{0009148}.}

Thus {\em if $K=T_p$} --- the tension of the original D-$p$ brane --- 
this would {\em prove}
the conjecture that the tachyon potential has an extremum where the
negative contribution due to the potential energy exactly cancels the
tension of the D-brane. 

We shall refer to the solution $(a=\infty, u^i=0)$ as the 
{\em vacuum} solution.

\item There is a solution where 
\be \label{e5}
u^i = \left\{
\begin{array}{ll}
\infty \quad &\mbox{for } 1\le i\le q,\\
0 \quad &\mbox{otherwise},
\end{array}\right. 
\ee
and $a$ determined as a function of the $u^i$'s\cite{0009148}.
This configuration describes a codimension $q$ soliton with energy per unit
$(p-q)$-volume, measured above the energy of the {\it vacuum
solution}, given by:
\be \label{e6}
\Delta{\cal E} = (2\pi)^{q}  K \, .
\ee
If $K=T_p$, this is precisely the correct formula for the tension of
the D-$(p-q)$-brane. 
\end{enumerate}

It is clear from above that in order to establish that the tachyon
dynamics in the background independent open string field theory reproduces
the conjectured relations involving tachyon condensation, we need to show
that $K=T_p$. This is what we shall prove in this paper. To this end, we
compute the on-shell three tachyon amplitude from the cubic open string
field theory\cite{SFT}, and compare it with the same amplitude 
computed in the background independent string field theory\cite{9208027}. 
(Notice that the three point tachyon amplitude calculated in 
Ref.\cite{0009148} vanishes on-shell. This does not agree with the
result of the first quantised theory.) 

\bigskip

Recall that the euclidean
action of the cubic open string field theory describing the dynamics of a
D-$p$-brane is given by
\be \label{e7}
S_{cubic}=2\pi^2 T_p \left( {1\over 2} \langle\Phi|Q_B|\Phi\rangle +
{1\over 3}
\langle f_1\circ\Phi(0) f_2\circ\Phi(0) f_3\circ\Phi(0)\rangle\right) \, ,
\ee
where $|\Phi\rangle$ is the string field represented by a ghost number 1
state in the Hilbert space of the first quantised theory, 
$f_i$'s are known conformal
maps reviewed in Ref.\cite{9911116}, and $f_i\circ\Phi(0)$ denotes the
conformal
transform of the vertex operator $\Phi(0)$ by $f_i$. The normalisation
factor $2\pi^2 T_p$ was derived in Ref.\cite{9911116}, and will be
important for
our analysis. The Fock vacuum $|k\rangle\equiv
e^{ik\cdot X(0)}|0\rangle$, where $|0\rangle$ is the SL(2,R) invariant
vacuum and $k^\mu$ labels momentum along the world-volume of the D-brane, 
is normalised as follows:
\be \label{e8}
\langle k| c_{-1} c_0 c_1|k'\rangle = (2\pi)^{p+1} \delta(k+k')\, .
\ee
We shall interpret $(2\pi)^{p+1}\delta(0)$ as the volume
of the D-$p$-brane world-volume.

Let us consider a tachyonic string field configuration of the form
\be \label{e9}
|\Phi\rangle = \int d^{p+1}k \, T(k) c_1 |k\rangle\, ,
\ee
with $T(k)$ supported over near on-shell momentum 
$k^2\simeq 1$.
Substituting \refb{e9} into \refb{e7}, and keeping only the leading
order terms in $(k^2-1)$ in both the quadratic and the cubic terms, we
arrive at the action:
\ben \label{e10}
S_{cubic} &\simeq& 2\pi^2 T_p \Bigg[{1\over 2} \int d^{p+1}k \int d^{p+1}k'
\, (2\pi)^{p+1} \delta(k+k') (k^2 - 1) T(k) T(k') \nonumber \\
&& +{1\over 3} 
\int d^{p+1}k \int d^{p+1}k' \int d^{p+1}k'' \, (2\pi)^{p+1}
\delta(k+k'+k'')
T(k) T(k') T(k'')\Bigg]\, .
\een
This encodes information about the on-shell three tachyon amplitude in the
cubic open string field theory.

\bigskip

Let us now turn to the background independent open string field theory. In
this case a near on-shell tachyon field configuration is represented by a
boundary perturbation of the form:
\be \label{e11}
\int {d\theta\over 2\pi} \VV(\theta) =
\int {d\theta\over 2\pi} \int d^{p+1} k \, \phi(k) e^{ik\cdot X(\theta)}\,
,
\ee
with $\phi(k)$ supported over near on-shell momentum 
$k^2\simeq 1$.
We shall use the normalisation \refb{e8} for computing
correlation functions in the world-sheet theory in the absence of 
any boundary perturbation. The action of the background independent 
open string field theory is given by eqn.\refb{e3}. 
To calculate the quadratic term it is sufficient to replace
$\langle~\rangle_\VV$ by 
$\langle~\rangle$ and use the relation 
\be \label{e12}
\{Q_B, c(\theta) e^{ik\cdot X(\theta)}\} = (k^2 -1) \p c(\theta) c(\theta)
e^{ik\cdot X(\theta)}\, .
\ee
Substituting this in eqn.\refb{e3}, and keeping terms to leading order in
$(k^2-1)$, we get the near on-shell quadratic term:
\be \label{e13}
S^{(2)}_{BI} =
{K\over 4} \int d^{p+1} k \int d^{p+1} k' \, (2\pi)^{p+1} \delta(k+k')
(k^2
-1)
\phi(k) \phi(k')\, .
\ee 

Next we evaluate the on-shell three tachyon coupling. Unfortunately, direct
determination of this coupling is difficult due to the problem with
ultraviolet divergences on the world-sheet\cite{9311177}, so we shall take
recourse to an indirect method.\footnote{Renormalization group equations
have been used in the past to derive on- and off-shell tachyon
amplitudes in open string theory\cite{RENORM}.} We use the fact that
whenever
the world-sheet action \refb{e1} describes a conformal field theory, the
corresponding string field configuration is a solution of the equations of
motion\cite{9208027}. Thus the equations of motion derived from the 
string field theory action $S_{BI}$
must be proportional to the $\beta$-functions of the boundary conformal
field theory described by the action \refb{e1}. Now if,
\be \label{e14}
\VV= \sum_{\alpha} \lambda^\alpha \VV_\alpha\, ,
\ee
where $\VV_\alpha$ are primary vertex operators of dimensions
$h_\alpha\simeq 1$,
with the operator product expansion 
\be \label{e15}
\VV_\alpha(x) \VV_\beta(y) \simeq {C_{\alpha\beta}^{\gamma}\over
|x-y|^{h_\alpha+h_\beta-h_\gamma}} \VV_\gamma(y)\, ,
\ee
the $\beta$-function associated with the coupling $\lambda^\alpha$ to
second order in $\lambda$ is given by\cite{BETA}
\be \label{e16}
\beta^\alpha(\lambda) \propto (h_\alpha -1) \lambda^\alpha + {1\over 2\pi} 
C^\alpha_{\beta\gamma}
\lambda^\beta\lambda^\gamma\, .
\ee
The factor of $(2\pi)^{-1}$ in front of the 
second term can be
traced to the normalisation factor of $(2\pi)^{-1}$ appearing in front of
the boundary perturbation in eqn.\refb{e1}. Since the operators
$e^{ik \cdot X}$
have conformal weights $h(k)=k^2$, and
satisfy the operator product expansion
\be \label{e17}
e^{ik\cdot X(x)} e^{ik'\cdot X(y)} \simeq 
{1\over |x-y|^{k^2 + k^{\prime 2} - (k+k')^2}}
e^{i(k+k')\cdot X(y)}\, ,
\ee
the equation of motion for the (near on-shell) tachyon field 
in BIOSFiT is:
\be \label{e18}
(k^2 -1) \phi(k) + {1\over 2\pi} \int d^{p+1}k' \int d^{p+1}k'' \, 
\delta(k -
k' -
k'') \phi(k') \phi(k'') = 0\, .
\ee
The cubic part of the background independent string field theory 
action can now be constructed, the normalisation being determined from 
eqn.\refb{e13}. Near on-shell the BIOSFiT action upto cubic order 
in $\phi(k)$ is given by,
\ben \label{e19}
S_{BI} &\simeq& {K\over 4} 
\Bigg[ \int d^{p+1} k \int d^{p+1} k' \, (2\pi)^{p+1}
\delta(k+k') (k^2 -1)
\phi(k) \phi(k') \nonumber \\
&& + {1\over 3\pi} 
\int d^{p+1}k \int d^{p+1}k' \int d^{p+1}k'' \, (2\pi)^{p+1}
\delta(k+k'+k'')
\phi(k) \phi(k') \phi(k'')\Bigg]\, .
\een
The above can be compared with the corresponding result of the cubic
open string field theory \refb{e10}. The quadratic terms in
the two actions agree under the identification:
\be \label{e20}
T(k) = {1\over 2\pi} \, \sqrt { K\over T_p} \, \phi(k) + \cdots\, ,
\ee
where the dots denote terms linear in $\phi(k)$ which vanish on-shell, and
terms quadratic and higher orders in $\phi(k)$.
This relates the tachyon fields in the two string field theories.
Requiring that the cubic terms match leads to
\be \label{e21}
K=T_p\, ,
\ee
the relation we set out to prove.
Thus the background independent string field theory
provides a verification of all the conjectures involving tachyon
condensation on the bosonic D-branes. It should be possible to generalise
this to prove the corresponding conjectures in superstring theory.

\medskip

{\bf Acknowledgement}: We would like to thank B.~Zwiebach for useful
discussions.

\end{document}